\documentclass[proceedings]{JHEP}
\def\pl#1#2#3{{\it Phys. Lett.}
{\bf #1~B}~(#2)~#3}
\def\apjl#1#2#3{{\it Astrophys. J. Lett.}
{\bf #1}~(#2)~#3}
\def\apj#1#2#3{{\it Astrophys. Journal}
{\bf #1}~(#2)~#3}
\def\jetpl#1#2#3{{\it JETP Lett.}
{\bf #1}~(#2)~#3}
\def\jetpsp#1#2#3{{\it JETP (Sov. Phys.)}
{\bf #1}~(#2)~#3}
\def\spss#1#2#3{{\it Sov. Phys. -Solid State}
{\bf #1}~(#2)~#3}
\def\jpa#1#2#3{{\it J. Phys.}
{\bf A~#1}~(#2)~#3}
\def\pr#1#2#3{{\it Phys. Reports}
{\bf #1}~(#2)~#3}
\def\mnras#1#2#3{{\it Mon. Not. Roy. Astr. Soc.}
{\bf #1}~(#2)~#3}
\def\n#1#2#3{{\it Nature}
{\bf #1}~(#2)~#3}
\def\cmp#1#2#3{{\it Commun. Math. Phys.}
{\bf #1}~(#2)~#3}
\def\prsla#1#2#3{{\it Proc. Roy. Soc. London}
{\bf A~#1}~(#2)~#3}
\def\ptp#1#2#3{{\it Prog. Theor. Phys.}
{\bf #1}~(#2)~#3}

\conference{Corfu Summer Institute on Elementary Particle
Physics, 1998}

\title{Introduction to Cosmology}

\author{G. Lazarides\\
Physics Division, School of Technology,\\
Aristotle University of Thessaloniki,\\
Thessaloniki 540 06, Greece\\
E-mail: \email{lazaride@eng.auth.gr}}

\abstract
{The standard big bang cosmological model and the history
of the early universe according to the grand unified
theories of strong, weak and electromagnetic interactions
are summarized. The shortcomings of big bang
are discussed together with their resolution by inflationary
cosmology. Inflation and the subsequent oscillation and decay
of the inflaton field are studied. The density perturbations
produced during inflation and their evolution during the
matter dominated era are analyzed. The temperature
fluctuations of the cosmic background radiation are summarized.
Finally, the nonsupersymmetric as well as the supersymmetric
hybrid inflationary model is described.}

\begin{document}

\section{The Big Bang Model}
\label{sec:bigbang}

\par
The discovery of the cosmic background radiation (CBR) in
1964 together with the observed Hubble expansion of the
universe had established hot big bang cosmology as a viable
model of the universe. The success of the theory of
nucleosynthesis in reproducing the observed abundance
pattern of light elements together with the proof of the
black body character of the CBR then established hot big
bang as the standard cosmological model. This model
combined with grand unified theories (GUTs) of strong,
weak and electromagnetic interactions provides an
appropriate framework for discussing the very early stages
of the universe evolution. A brief introduction to hot big
bang follows.

\subsection{Hubble Expansion}
\label{subsec:hubble}

\par
For cosmic times $t\stackrel{_{>}}{_{\sim }} t_{P}
\equiv M_{P}^{-1}\sim 10^{-44}~{\rm{sec}}$ ($M_{P}=
1.22\times 10^{19}~{\rm{GeV}}$ is the Planck scale)
after the big bang, quantum fluctuations of gravity cease
to exist. Gravitation can then be adequately described by
classical relativity. Strong, weak and electromagnetic
interactions, however, require relativistic quantum field
theoretic treatment and are described by gauge theories.

\par
An important principle, on which the standard
big bang (SBB) cosmological model \cite{wkt} is based,
is that the universe is homogeneous and isotropic. The
strongest evidence so far for this {\it cosmological
principle} is the observed \cite{cobe} isotropy
of the CBR. Under this assumption, the four
dimensional spacetime in the universe is described
by the Robertson-Walker metric
$$
ds^{2}=-dt^{2}+
$$
\begin{equation}
a^{2}(t)\left[\frac{dr^{2}}{1-kr^2}
+r^{2}(d\theta^{2}+\sin^{2}\theta~d\varphi^{2})
\right]~,
\label{eq:rw}
\end{equation}
where $r$, $\varphi$ and $\theta$ are `comoving' polar
coordinates, which remain fixed for objects that have no
other motion than the general expansion of the universe.
The parameter $k$ is the `scalar curvature' of the 3-space
and $k=0$, $k>0$ or $k<0$ correspond to flat, closed or
open universe. The dimensionless parameter $a(t)$ is the
`scale factor' of the universe and describes cosmological
expansion. We normalize it by taking $a_{0}\equiv
a(t_{0})=0$, where $t_{0}$ is the present cosmic time.

\par
The `instantaneous' radial physical distance is given by
\begin{equation}
R=a(t)\int_{0}^{r}\frac{dr}{(1-kr^{2})^{1/2}}~\cdot
\label{eq:dist}
\end{equation}
For flat universe ($k=0$), $\bar{R}=a(t)\bar{r}$
($\bar{r}$ is a `comoving' and  $\bar{R}$ a physical
vector in 3-space) and the velocity of an object is
\begin{equation}
\bar{V}=\frac{d\bar{R}}{dt}=\frac{\dot{a}}{a}\bar{R}
+a\frac{d\bar{r}}{dt}~,
\label{eq:velocity}
\end{equation}
where overdots denote derivation with respect to cosmic
time. The second term in the right hand side (rhs) of
this equation is the `peculiar velocity', $\bar{v}=a(t)
\dot{\bar{r}}$, of the object, i.e., its velocity with
respect to the `comoving' coordinate system. For
$\bar{v}=0$, Eq.(\ref{eq:velocity}) becomes
\begin{equation}
\bar{V}=\frac{\dot{a}}{a}\bar{R}\equiv H(t)\bar{R}~,
\label{eq:hubblelaw}
\end{equation}
where $H(t)\equiv \dot{a}(t)/a(t)$ is the Hubble
parameter. This is the well-known Hubble law asserting
that all objects run away from each other
with velocities proportional to their distances and is
considered as the first success of SBB cosmology.

\subsection{Friedmann Equation}
\label{subsec:friedmann}

\par
Homogeneity and isotropy of the universe imply that the
energy momentum tensor takes the diagonal form
$(T_{\mu}^{\nu})= {\rm{diag}}(-\rho, p, p, p)$,
where $\rho$ is the energy density of the universe and
$p$ the pressure. Energy momentum conservation
(${T_{\mu~;\nu}^{~\nu}}=0$) then takes the form of the
continuity equation
\begin{equation}
\frac{d\rho}{dt}=-3H(t)(\rho+p)~,
\label{eq:continuity}
\end{equation}
where the first term in the rhs describes the dilution of
the energy due to the expansion of the universe and the
second term corresponds to the work done by pressure.
Eq.(\ref{eq:continuity}) can be given
the following more transparent form
\begin{equation}
d\left(\frac{4\pi}{3}a^{3}\rho\right)=
-p~4\pi a^{2}da~,
\label{eq:cont}
\end{equation}
which indicates that the energy loss of a `comoving'
sphere of radius $\propto a(t)$ equals the work
done by pressure on its boundary as it expands.

\par
For a universe described by the Robertson-Walker metric in
Eq.(\ref{eq:rw}), Einstein's equations
\begin{equation}
R_{\mu}^{~\nu}-\frac{1}{2}~\delta_{\mu}^{~\nu}R=
8\pi G~T_{\mu}^{~\nu}~,
\label{eq:einstein}
\end{equation}
where $R_{\mu}^{~\nu}$ and $R$ are the Ricci tensor
and scalar curvature tensor and $G\equiv
M_{P}^{-2}$ is the Newton's constant, lead to the
Friedmann equation
\begin{equation}
H^{2}\equiv \left(\frac{\dot{a}(t)}{a(t)}
\right)^{2}=\frac{8\pi G}{3}\rho-\frac{k}{a^{2}}~\cdot
\label{eq:friedmann}
\end{equation}

\par
Averaging $p$, we write $\rho+p=\gamma \rho$.
Eq.(\ref{eq:continuity}) then becomes $\dot{\rho}=
-3H\gamma\rho$, which gives $d \rho/\rho=-3 \gamma da/a$
and $\rho \propto a^{-3 \gamma}$. For a universe
dominated by pressureless matter, $p=0$ and, thus,
$\gamma=1$, which gives $\rho\propto a^{-3}$. This is
easily interpreted as mere dilution of a fixed number
of particles in a `comoving' volume due to the
cosmological expansion. For a radiation dominated universe,
$p=\rho/3$ and, thus, $\gamma=4/3$, which
gives $\rho\propto a^{-4}$. In this case, we get an extra
factor of $a(t)$ due to the red-shifting of all wave-lengths
by the expansion. Substituting $\rho \propto a^{-3 \gamma}$
in Friedmann equation with $k=0$, we get $\dot{a}/a \propto
a^{-3 \gamma/2}$ and, thus, $a(t)\propto t^{2/3\gamma}$.
Taking into account the normalization of $a(t)$
($a(t_{0})=1$), this gives
\begin{equation}
a(t)=(t/t_{0})^{2/3\gamma}~.
\label{eq:expan}
\end{equation}
For a matter dominated universe, we get the expansion law
$a(t)=(t/t_{0})^{2/3}$. `Radiation', however, expands as
$a(t)=(t/t_{0})^{1/2}$.

\par
The universe in its early stages of evolution is radiation
dominated and its energy density is
\begin{equation}
\rho=\frac{\pi^{2}}{30}\left(N_{b}+\frac{7}{8}N_{f}
\right)T^{4}\equiv c~T^{4}~,
\label{eq:boltzman}
\end{equation}
where $T$ is the cosmic temperature and $N_{b}$ ($N_{f}$)
is the number of massless bosonic (fermionic) degrees of
freedom. The combination $g_{*}=N_{b}+(7/8)N_{f}$ is
called effective number of massless degrees of freedom.
The entropy density is
\begin{equation}
s= \frac{2\pi^{2}}{45}~g_{*}~T^{3}~.
\label{eq:entropy}
\end{equation}
Assuming adiabatic universe evolution, i.e., constant
entropy in a `comoving' volume ($sa^{3}={\rm{constant}}$),
we obtain the relation $aT={\rm{constant}}$. The
temperature-time relation during radiation dominance is then
derived from Friedmann equation (with $k=0$):
\begin{equation}
T^{2}=\frac{M_{P}}{2(8\pi c/3)^{1/2}t}~\cdot
\label{eq:temptime}
\end{equation}
We see that classically the expansion starts at $t=0$ with
$T=\infty$ and $a=0$. This initial singularity is, however,
not physical since general relativity fails at cosmic times
smaller than about the Planck time $t_{P}$. The only
meaningful statement is that the universe, after a yet
unknown initial stage, emerges at a cosmic time
$\sim t_{P}$ with temperature $T\sim M_{P}$.

\subsection{Important Cosmological Parameters}
\label{subsec:parameter}

\par
The most important parameters describing the expanding
universe are the following:
\begin{list}
\setlength{\rightmargin=0cm}{\leftmargin=0cm}
\item[{\bf i.}] The present value of the Hubble parameter
(known as Hubble constant) $H_{0}\equiv
H(t_{0})=100~h~\rm{km}~\rm{sec}^{-1}~\rm{Mpc}^{-1}$
($0.4\stackrel{_{<}}{_{\sim }}h
\stackrel{_{<}}{_{\sim }}0.8$).
\item[{\bf ii.}]The fraction $\Omega=\rho/\rho_{c}$,
where $\rho_{c}$ is the critical density corresponding
to a flat universe ($k=0$). From Friedmann equation,
$\rho_{c}=3H^{2}/8\pi G$ and, thus,
$\Omega=1+k/a^{2}H^{2}$.
$\Omega=1$, $\Omega>1$ or $\Omega<1$ correspond to
flat, closed or open universe. Assuming inflation (see
below), the present value of $\Omega$ must be
$\Omega_{0}=1$. However, the baryonic contribution
to $\Omega$ is $\Omega_{B}\approx 0.05-0.1$
\cite{deuterium}. This indicates that most of the energy
in the universe must be in nonbaryonic form.
\item[{\bf iii.}]The deceleration parameter
\begin{equation}
q=-\frac{(\ddot{a}/\dot{a})}{(\dot{a}/a)}
=\frac{\rho+3p}{2\rho_{c}}~\cdot
\label{eq:decel}
\end{equation}
For `matter', $q=\Omega/2$ and, thus, inflation implies
that the present deceleration parameter is $q_{0}=1/2$.
\end{list}

\subsection{Particle Horizon}
\label{subsec:parhor}

\par
Light travels only a finite distance from the time
of big bang ($t=0$) till some cosmic time $t$. From the
Robertson-Walker metric in Eq.(\ref{eq:rw}), we find that
the propagation of light along the radial direction is
described by the equation $a(t)dr=dt$. The particle
horizon, which is the `instantaneous' distance at time $t$
travelled by light since the beginning of time, is then
given by
\begin{equation}
d_{H}(t)=a(t)\int_{0}^{t}\frac{dt^{\prime}}
{a(t^{\prime})}~\cdot
\label{eq:hor}
\end{equation}
The particle horizon is a very important notion since it
coincides with the size of the universe
already seen at time $t$ or, equivalently, with the
distance at which causal contact has been established
at $t$. Eqs.(\ref{eq:expan}) and (\ref{eq:hor}) give
\begin{equation}
d_{H}(t)=\frac{3\gamma}{3\gamma-2}t~,
~\gamma\neq 2/3~.
\label{eq:hort}
\end{equation}
Also,
\begin{equation}
H(t)=\frac{2}{3\gamma}t^{-1}~,
~d_{H}(t)=\frac{2}{3\gamma-2}H^{-1}(t)~.
\label{eq:hubblet}
\end{equation}
For `matter' (`radiation'), this becomes $d_{H}(t)=3t=
2H^{-1}(t)$ ($d_{H}(t)=2t=H^{-1}(t)$). The present
particle horizon is $d_{H}(t_{0})=2H_{0}^{-1}\approx
6,000~h^{-1}~{\rm{Mpc}}$, the present cosmic time
is $t_{0}=2H_{0}^{-1}/3\approx 6.7\times 10^{9}
~h^{-1}~{\rm{years}}$ and the present value of the
critical density
is $\rho_{c}=3H_{0}^{2}/8\pi G\approx 1.9\times
10^{-29}~h^{-2}~{\rm{gm/cm^{3}}}$.

\subsection{Brief History of the Early Universe}
\label{subsec:history}

\par
We will now briefly describe the early stages of
the universe evolution according to GUTs \cite{ggps}.
We will take a GUT based on the gauge group $G$ ($=SU(5)$,
$SO(10)$, $SU(3)^{3}$, ...) with or without supersymmetry.
At a superheavy scale $M_{X}\sim 10^{16}~{\rm{GeV}}$
(the GUT mass scale), $G$ breaks to the standard model
gauge group $G_{S}=
SU(3)_{c}\times SU(2)_{L}\times U(1)_{Y}$ by the
vacuum expectation value (vev) of an appropriate higgs
field $\phi$. (For simplicity, we will consider that
this breaking occurs in just one step.) $G_{S}$ is,
subsequently, broken to $SU(3)_{c}\times U(1)_{em}$
at the electroweak scale $M_{W}$.

\par
GUTs together with the SBB cosmological model (based on
classical gravitation) provide a suitable framework for
discussing the early history of the universe for cosmic
times $\stackrel{_{>}}{_{\sim }} 10^{-44}~
{\rm{sec}}$. They predict that the universe, as it
expands and cools down after the big bang, undergoes
\cite{kl} a series of phase transitions during which the
initial gauge symmetry is gradually reduced and several
important phenomena take place.

\par
After the big bang, the GUT gauge group $G$ was unbroken
and the universe was filled with a hot `soup' of massless
particles which included not only photons, quarks, leptons
and gluons but also the weak gauge boson $W^{\pm}$, $Z^{0}$,
the GUT gauge bosons $X$, $Y$, ... as well as several higgs
bosons. (In the supersymmetric case, all the supersymmetric
partners of these particles were also present.) At cosmic
time $t\sim 10^{-37}~{\rm{sec}}$ corresponding to
temperature $T\sim 10^{16}~{\rm{GeV}}$, $G$ broke down to
$G_{S}$ and the $X$, $Y$, ... gauge bosons together with
some higgs bosons acquired superheavy masses of order $M_{X}$.
The out-of-equilibrium decay of these superheavy particles
can produce \cite{bau} the observed baryon asymmetry of the
universe (BAU). Important ingredients for this mechanism to
work are the violation of baryon number, which is inherent
in GUTs, and C and CP violation. This is the second
important success of the SBB model.

\par
During the GUT phase transition, topologically stable
extended objects
\cite{kibble} such as magnetic monopoles \cite{monopole},
cosmic strings \cite{string} or domain walls \cite{wall}
can also be produced. Monopoles, which exist in all GUTs,
can lead into cosmological problems \cite{preskill} which
are, however, avoided by inflation \cite{guth,lindebook}
(see Secs.\ref{subsec:monopole} and \ref{subsec:infmono}).
This is a period of an exponentially fast expansion of the
universe which can occur during some GUT phase transition.
Strings can contribute \cite{zel} to the primordial
density fluctuations necessary for structure formation
\cite{structure} in the universe whereas domain walls
are \cite{wall} absolutely catastrophic and GUTs
predicting them should be avoided or inflation should be
used to remove them from the scene.

\par
At $t\sim 10^{-10}~{\rm{sec}}$ or $T\sim 100~{\rm{GeV}}$,
the electroweak transition takes place and $G_{S}$ breaks
to $SU(3)_{c}\times U(1)_{em}$. The $W^{\pm}$,
$Z^{0}$ gauge bosons together with the electroweak higgs
fields acquire masses $\sim M_{W}$. Subsequently, at
$t\sim 10^{-4}~{\rm{sec}}$ or $T\sim 1~{\rm{GeV}}$,
color confinement sets in and the quarks get bounded
forming hadrons.

\par
The direct involvement of particle physics essentially ends
here since most of the subsequent phenomena fall into the
realm of other branches. We will, however, sketch
some of them since they are crucial for understanding the
earlier stages of the universe evolution where their origin
lies.

\par
At $t\approx 180~{\rm{sec}}$ ($T\approx 1~{\rm{MeV}}$),
nucleosynthesis takes place, i.e., protons and neutrons
form nuclei. The abundance of light elements ($D$,
$^{3}He$, $^{4}He$ and $^{7}Li$) depends
\cite{peebles} crucially on the number of light particles
(with mass $\stackrel{_{<}}{_{\sim }} 1~{\rm{MeV}}$),
i.e., the number of light neutrinos, $N_{\nu}$, and
$\Omega_{B}h^{2}$. Agreement with observations
\cite{deuterium} is achieved for $N_{\nu}=3$ and
$\Omega_{B}h^{2}\approx 0.019$. This is
the third success of SBB cosmology. Much later, at the so
called `equidensity' point, $t_{\rm{eq}}\approx 3,000~
{\rm{years}}$, `matter' dominates over `radiation'.

\par
At cosmic time $t\approx 200,000~h^{-1} {\rm{years}}$
($T\approx 3,000~{\rm{K}}$), we have the `decoupling' of
`matter' and `radiation' and the `recombination' of atoms.
After this, `radiation' evolves as an independent (not
interacting) component of the universe and is detected
today as CBR with temperature $T_{0}\approx 2.73~{\rm{K}}$.
The existence of this radiation is the fourth
important success of the theory of big bang. Finally,
structure formation \cite{structure} in the universe
starts at $t\approx 2\times 10^{8}~{\rm{years}}$.

\section{Shortcomings of Big Bang}
\label{sec:short}

The SBB cosmological model has been very successful in
explaining, among other things, the Hubble expansion of the
universe, the existence of the CBR and the abundances of the
light elements which were formed during primordial
nucleosynthesis. Despite its great successes, this model had
a number of long-standing shortcomings which we will now
summarize:

\subsection{Horizon Problem}
\label{subsec:horizon}

The CBR, which we receive now, was emitted at the time of
`decoupling' of matter and radiation when the
cosmic temperature was $T_d  \approx 3,000~\rm{K}$. The
decoupling time, $t_d$, can be calculated from
\begin{equation}
\frac {T_0}{T_d} =
\frac {2.73~\rm{K}}{3,000~\rm{K}} =
\frac {a (t_d)}{a(t_0)} =
\left(\frac {t_d}{t_0}\right)^{2/3}\cdot
\label{eq:dec}
\end{equation}
It turns out that $t_d \approx 200,000~h^{-1}$ years.

\par
The distance over which the photons of the CBR have
travelled since their emission is
$$
a(t_0) \int^{t_{0}} _{t_{d}}
\frac {dt^\prime}{a(t^\prime)} = 3t_0
\left[1 - \left(\frac {t_d}{t_0}\right)^{2/3}\right]
$$
\begin{equation}
\approx 3t_0 \approx 6,000~h^{-1}~\rm{Mpc}~,
\label{eq:lss}
\end{equation}
which essentially coincides with the present particle
horizon size. A sphere around us with radius equal to this
distance is called the `last scattering surface' since the
CBR observed now has been emitted from it. The particle
horizon size at $t_d$ was $2H^{-1} (t_d) = 3t_d \approx
0.168~h^{-1}~
\rm{Mpc}$ and expanded till the present time to become
equal to $0.168~h^{-1} (a(t_0)/a(t_d))~{\rm{Mpc}}
\approx 184~h^{-1}$ Mpc. The angle subtended by this
`decoupling' horizon at present is $\theta_{d} \approx
184/6,000 \approx 0.03~\rm{rads} \approx 2~^o$.
Thus, the sky splits into $4 \pi/(0.03)^2 \approx 14,000$
patches that never communicated causally before sending
light to us. The question then arises how come the
temperature of the black body radiation from all these
patches is so accurately tuned as the measurements of the
cosmic background explorer \cite{cobe} (COBE) require
($\delta T/T \approx 6.6 \times 10^{-6}$).

\subsection{Flatness Problem}
\label{subsec:flatness}

The present energy density, $\rho$, of the universe has
been observed to lie in the relatively narrow range
$0.1 \rho_c \stackrel{_{<}}
{_{\sim }}\rho \stackrel{_{<}}{_{\sim }}2 \rho_c$,
where $\rho_c$ is the critical energy density corresponding
to a flat universe. The lower bound has been derived from
estimates of galactic masses using the virial theorem whereas
the upper bound from the volume expansion rate implied by the
behavior of galactic number density at large distances.
Eq.(\ref{eq:friedmann}) implies that
$(\rho - \rho_c)/\rho_c =3 (8 \pi G \rho_c)^{-1}
(k/a^2)$ is proportional to $a$, for matter dominated
universe. Consequently, in the early universe, we have
$ |(\rho - \rho_c)/\rho_c|\ll 1$ and the question  arises
why the initial energy density of the universe was so finely
tuned to be equal to its critical value.

\subsection{Magnetic Monopole Problem}
\label {subsec:monopole}

This problem arises only if we combine the SBB model with
GUTs \cite{ggps} of strong, weak and electromagnetic
interactions. As already indicated, according to GUTs, the
universe underwent \cite{kl} a phase transition during
which the GUT gauge symmetry group, $G$, broke to $G_{S}$.
This breaking was
due to the fact that, at a critical temperature $T_c$, an
appropriate higgs field, $\phi$, developed a nonzero vev.
Assuming that this phase transition was a second order one,
we have $\langle \phi\rangle(T) \approx \langle \phi
\rangle(T=0)( 1 - T^2/T^2_c)^{1/2}$, $m_H (T)\approx
\lambda \langle \phi\rangle (T)$, for the temperature
dependent vev and mass of the higgs field respectively at
$T \leq T_c$ ($\lambda$ is an appropriate higgs coupling
constant).

\par
The GUT phase transition produces magnetic monopoles
\cite{monopole} which are localized deviations from the
vacuum with radius $\sim M_X^{-1}$, energy $\sim M_X/
\alpha_G$ and $\phi =0$ at their center ($\alpha_G=
g^2_{G}/4\pi$ with $g_G$ being the GUT gauge coupling
constant). The vev of
the higgs field on a sphere, $S^2$, with radius
$\gg M_{X}^{-1}$ around the monopole lies on the vacuum
manifold $G/G_S$ and we, thus, obtain a  mapping:
$S^2\longrightarrow G/G_S$. If this mapping is
homotopically nontrivial the topological stability of the
monopole is guaranteed.

\par
Monopoles can be produced when the fluctuations of $\phi$
over $\phi=0$ between the vacua at
$\pm \langle \phi\rangle(T)$ cease to be frequent.
This takes place when the free energy needed for $\phi$ to
fluctuate from $\langle \phi\rangle (T)$ to zero in a
region of radius equal to the higgs correlation length
$\xi(T) = m^{-1}_H (T)$  exceeds $T$. This condition
reads $(4\pi/3) \xi^3 \Delta V \stackrel{_{>}}
{_{\sim }} T$, where $\Delta V \sim \lambda^2
\langle \phi\rangle^4$ is the difference in free energy
density between $\phi =0$ and $\phi=
\langle \phi\rangle(T)$. The Ginzburg temperature
\cite{ginzburg}, $T_G$, corresponds to the saturation
of this inequality. So, at
$T \stackrel{_{<}}{_{\sim }} T_G$, the fluctuations
over $\phi=0$ stop and $\langle \phi\rangle$ settles
on the vacuum manifold $G/G_S$. At $T_G$, the universe
splits into regions of size
$\xi_G \sim (\lambda^2 T_c)^{-1}$,
the higgs correlation length at $T_G$, with the
higgs field being more or less aligned in each region.
Monopoles are produced at the corners where such regions
meet (Kibble \cite{kibble} mechanism) and their number
density is estimated to be $n_M \sim {\rm{p}}
\xi_{G}^{-3} \sim {\rm{p}} \lambda^{6} T_{c}^{3}$,
where $\rm{p} \sim \rm{1/10}$ is a geometric factor.
The `relative' monopole number density then turns out to
be $r_M =n_M/T^3 \sim \rm{10^{-6}}$. We can derive a
lower bound on $r_M$ by employing causality. The higgs
field $\phi$ cannot be correlated at distances bigger
than the particle horizon size, $2t_G$, at $T_G$. This
gives the causality bound
\begin{equation}
n_M  \stackrel{_{>}}{_{\sim }}\frac {\rm{p}}
{\frac{4 \pi}{3}(2t_G)^3}~,
\label{eq:causal}
\end{equation}
which implies that $r_M\stackrel{_{>}}{_{\sim }}
\rm{10^{-10}}$.

\par
The subsequent evolution of monopoles, after $T_G$, is
governed by the equation \cite{preskill}
\begin{equation}
\frac {dn_M}{dt} =
- D n_{M}^{2} - 3 \frac {\dot{a}}{a} n_{M}~,
\label{eq:evol}
\end{equation}
where the first term in the rhs (with $D$
being an appropriate constant) describes the dilution of
monopoles due to their annihilation with antimonopoles
while the second term corresponds to their dilution by
the general cosmological expansion. The monopole-antimonopole
annihilation proceeds as follows. Monopoles diffuse towards
antimonopoles in the plasma of charged particles, capture
each other in Bohr orbits and eventually annihilate. The
annihilation is effective provided the mean free path of
monopoles in the plasma of charged particles does not
exceed their capture distance. This happens at cosmic
temperatures $T \stackrel{_{>}}{_{\sim }}
\rm{10^{12}~GeV}$. The overall result is that, if the
initial relative magnetic monopole density
$r_{M,\rm{in}} \stackrel{_{>}}{_{\sim }}\rm{10^{-9}}
( \stackrel{_{<}}{_{\sim }}\rm{10^{-9}})$, the final
one $r_{M,\rm{fin}} \sim 10^{-9} ( \sim r_{M,\rm{in}})$.
This combined with the causality bound yields
$r_{M,\rm{fin}} \stackrel{_{>}}{_{\sim }}
\rm{10^{-10}}$. However, the requirement that monopoles
do not dominate the energy density of the universe at
nucleosynthesis gives
\begin{equation}
r_M (T \approx 1~\rm{MeV}) \stackrel{_{<}}
{_{\sim }}\rm{10^{-19}}~,
\label{eq:nucleo}
\end{equation}
and we obtain a clear discrepancy of about ten orders
of magnitude.

\subsection{Density Fluctuations}
\label{subsec:fluct}

For structure formation~\cite{structure} in the universe,
we need a primordial density perturbation,
$\delta \rho/ \rho$, at all length scales with a nearly
flat spectrum \cite{hz}. We also need some explanation
of the temperature fluctuations, $\delta T/T$, of CBR
observed by COBE \cite{cobe} at angles
$\theta \stackrel{_{>}}{_{\sim }}
\theta_d \approx 2~^o$ which violate causality
(see Sec.\ref{subsec:horizon}).

\par
Let us expand $\delta \rho/\rho$ in plane waves
\begin{equation}
\frac {\delta \rho} {\rho} (\bar{r},t) =
\int d^3 k\delta_{\bar{k}}(t)e^{i\bar{k} \bar{r}}~,
\label{eq:plane}
\end{equation}
where $\bar{r}$ is a `comoving' vector in 3-space and
$\bar{k}$ is the `comoving' wave vector with
$k=|\bar{k}|$ being the `comoving' wave number
($\lambda=2 \pi/k$ is the `comoving' wave length whereas
the physical wave length is $ \lambda _{\rm{phys}}=
a(t) \lambda$). For $\lambda_{\rm{phys}} \leq H^{-1}$,
the time evolution of $\delta_{\bar{k}}$ is described by
the Newtonian equation
\begin{equation}
\ddot{\delta}_{\bar{k}} + 2 H \dot {\delta}_{\bar{k}}
+ \frac {v_{s}^{2}k^2}{a^2}\delta_{\bar{k}}=
4 \pi G \rho \delta_{\bar{k}}~,
\label{eq:newton}
\end{equation}
where the second term in the left hand side (lhs) comes
from the cosmological expansion and the third is the
`pressure' term ($v_s$ is the velocity of sound given by
$v^{2}_{s}=dp/d\rho$, where $p$ is the mean pressure).
The rhs of this equation corresponds to the gravitational
attraction.

\par
For the moment, let us put $H$=0 (static universe). In this
case, there exists a characteristic wave number $k_J$, the
Jeans wave number, given by $k^{2}_{J}=4 \pi G a^{2} \rho/
v^{2}_{s}$ and having the following property. For
$k \geq k_J$,  pressure dominates over gravitational
attraction and the density perturbations just oscillate,
whereas, for $k \leq k_J$, gravitational attraction
dominates and the density perturbations grow exponentially.
In particular, for $p$=0 (matter domination), $v_s=0$ and
all scales are Jeans unstable with
\begin{equation}
\delta_{\bar{k}} \propto {\rm{exp}}(t/\tau)~,~\tau=
(4 \pi G \rho)^{-1/2}~.
\label{eq:jeans}
\end{equation}

\par
Now let us take $H\neq 0$. Since the cosmological expansion
pulls the particles apart, we get a smaller growth:
\begin{equation}
\delta_{\bar{k}}\propto a(t) \propto t^{2/3}~,
\label{eq:growth}
\end{equation}
in the matter dominated case. For a radiation dominated
universe ($p \neq 0$), we get essentially no growth of
the density perturbations. This means that, in order to
have structure formation in the universe, which requires
$\delta \rho/\rho \sim 1$, we must have
\begin{equation}
(\frac {\delta \rho} {\rho})_{\rm{eq}}\sim
4 \times 10^{-5} ( \Omega_0 h)^{-2}~,
\label{eq:equi}
\end{equation}
at the `equidensity' point (where the energy densities of
matter and radiation coincide), since the available growth
factor for perturbations is given by $a_0/a_{\rm{eq}}
\sim 2.5 \times 10^4 (\Omega_0 h)^2$. Here $\Omega_0=
\rho_0/\rho_c$, where $\rho_0$ is the present energy
density of the universe. The question then is where these
primordial density fluctuations originate from.

\section{Inflation}
\label{sec:inflation}

Inflation~\cite{guth,lindebook} is an idea which solves
simultaneously all four cosmological puzzles and can be
summarized as follows. Suppose there is a real scalar
field $\phi$ (the inflaton) with (symmetric) potential
energy density $V(\phi)$ which is quite `flat' near
$\phi=0$ and has minima at $\phi =
\pm\langle \phi\rangle$ with
$V(\pm \langle \phi\rangle)=0$.
At high enough $T$'s, $\phi =0$ in the universe due to
the temperature corrections in $V(\phi)$. As $T$ drops,
the effective potential density approaches the $T$=0
potential but a little potential barrier separating
the local minimum at $\phi=0$ and the vacua at
$\phi = \pm\langle \phi\rangle$ still remains.
At some point, $\phi$ tunnels out to
$\phi_1 \ll\langle \phi\rangle$ and a bubble with
$\phi=\phi_1$ is created in the universe. The field
then rolls over to the minimum of $V(\phi)$ very slowly
(due to the flatness of the potential). During this slow
roll over, the energy density
$\rho \approx V(\phi=0) \equiv V_0$ remains essentially
constant for quite some time. The Lagrangian density
\begin{equation}
L=\frac{1}{2} \partial_{\mu} \phi \partial^{\mu} \phi
- V(\phi)
\label{eq:lagrange}
\end{equation}
gives the energy momentum tensor
\begin{equation}
T_{\mu}^{~\nu} = - \partial_\mu \phi \partial^\nu \phi +
\delta_{\mu}^{~\nu}\left(\frac{1}{2}
\partial_\lambda \phi \partial^\lambda \phi -
V (\phi)\right)~,
\label{eq:energymom}
\end{equation}
which during the slow roll over takes the form
$T_{\mu}^{~\nu} \approx - V_{0}~\delta_{\mu}^{~\nu}$.
This means that $\rho \approx -p \approx V_0$, i.e.,
the pressure $p$ is negative and equal in magnitude with the
energy density $\rho$, which is consistent with
Eq.(\ref{eq:continuity}). Since, as we will see, $a(t)$ grows
very fast, the `curvature' term, $k/a^2$,
in Eq.(\ref{eq:friedmann}) becomes subdominant and we get
\begin{equation}
H^2 \equiv \left(\frac {\dot{a}}{a}\right)^2 =
\frac {8 \pi G} {3} V_0~,
\label{eq:inf}
\end{equation}
which gives $a(t) \propto e^{Ht},~H^2 =(8 \pi G/3)V_0$ =
~constant. So the bubble expands exponentially for some time
and $a(t)$ grows by a factor
\begin{equation}
\frac {a(t_f)}{a(t_i)} ={\rm{exp}} H(t_f - t_i)
\equiv {\rm{exp}} H \tau~,
\label{eq:efold}
\end{equation}
between an initial ($t_i$) and a final ($t_f$) time.

\par
The inflationary scenario just described here, known as
new \cite{new} inflation (with the inflaton field starting
from the origin, $\phi$=0), is certainly not the only
realization of the idea of inflation. Another interesting
possibility is to consider the universe as it emerges at
the Planck time $t_{P}$, where the fluctuations of gravity
cease to exist. We can imagine a region of size
$\ell_{P} \sim M_{P}^{-1}$ where the inflaton
field acquires a
large and almost uniform value and carries negligible kinetic
energy. Under certain circumstances this region can inflate
(exponentially expand) as $\phi$ rolls down towards its
vacuum value. This type of inflation with the inflaton
starting from large values is known as the chaotic
\cite{chaotic} inflationary scenario.

\par
We will now show that, with an adequate number of e-foldings,
$N=H \tau$, the first three cosmological puzzles are easily
resolved (we leave the question of density perturbations for
later).

\subsection{Resolution of the Horizon Problem}
\label{subsec:infhor}

The particle horizon during inflation (exponential expansion)
\begin{equation}
d(t) = e^{Ht} \int^t_{t_{i}}
\frac {d t^\prime}{e^{Ht^\prime}}
\approx H^{-1}{\rm{exp}}H(t-t_i)~,
\label{eq:horizon}
\end{equation}
for $t-t_i \gg H^{-1}$, grows as fast as $a(t)$. At the
end of inflation ($t=t_f$),~
$d(t_f) \approx H^{-1}{\rm{exp}} H \tau$ and the field
$\phi$ starts oscillating about the
minimun of the potential at $\phi = \langle \phi\rangle$.
It then decays and `reheats' \cite{reheat} the universe at a
temperature $T_r \sim 10^9~{\rm{GeV}}$ \cite{gravitino}.
The universe, after that, goes back to normal big bang
cosmology. The horizon $d(t_{f})$ is stretched during the
period of $\phi$-oscillations by some factor $\sim 10^9$
depending on  details and between $T_r$ and the present era
by a factor $T_r/T_0$. So it finally becomes equal to
$H^{-1} e^{H \tau} 10^9 (T_r/T_0)$, which should exceed
$2H_{0}^{-1}$ in order to solve the horizon problem. Taking
$V_0 \approx M_{X}^{4},~M_{X} \sim 10^{16}$ GeV, we see
that, with $N = H \tau\stackrel{_{>}}{_{\sim }} 55$,
the horizon problem is evaded.

\subsection{Resolution of the Flatness Problem}
\label{subsec:infflat}

The `curvature' term of the Friedmann equation, at present,
is given by
\begin{equation}
\frac {k}{a^2} \approx \left(\frac {k}{a^2}\right)_{bi}
e^{-2H \tau}~10^{-18} \left(\frac {10^{-13}~{\rm{GeV}}}
{10^9~{\rm{ GeV}}}\right)^2,
\label{eq:curvature}
\end{equation}
where the terms in the rhs correspond to the `curvature'
term before inflation, and its growth factors during
inflation, during $\phi$-oscillations and after `reheating'
respectively. Assuming
$(k/a^2)_{bi} \sim (8 \pi G/3) \rho \sim H ^2$~~
$(\rho \approx V_0)$, we get $k/a_{0}^{2} H_{0}^{2}
\sim 10^{48}~e^{-2H \tau}$ which gives
$(\rho_0 - \rho_c)/\rho_c \equiv \Omega_0 - 1 =
k/a_{0}^{2} H_{0}^{2} \ll 1$, for
$H \tau \gg 55$. In fact, strong inflation implies that
the present universe is flat with a great accuracy.

\subsection{Resolution of the Monopole Problem}
\label{subsec:infmono}

It is obvious that, with a number of e-foldings
$\stackrel{_{>}}{_{\sim }} 55$, the primordial monopole
density is diluted by at
least 70 orders of magnitude and they become totally irrelevant.
Also, since $T_r \ll m_M$ (=the monopole mass), there is no
production of magnetic monopoles after `reheating'.

\section{Detailed Analysis of Inflation}
\label{sec:detail}

The Hubble parameter is not exactly constant during
inflation as we, naively, assumed so far. It actually
depends on the value of $\phi$:
\begin{equation}
H^{2}(\phi) = \frac {8 \pi G} {3} V (\phi)~.
\label{eq:hubble}
\end{equation}
To find the evolution equation for $\phi$ during inflation,
we vary the action
\begin{equation}
\int \sqrt{-{\rm{det}}(g)}~ d^{4}x \left(\frac {1}{2}
\partial_ {\mu} \phi \partial^{\mu} \phi - V(\phi) +
M(\phi)\right)~,
\label{eq:action}
\end{equation}
where $g$ is the metric tensor and $M(\phi)$ represents
the coupling of $\phi$ to `light' matter causing its decay.
We find
\begin{equation}
\ddot{\phi} + 3H \dot{\phi} + \Gamma_{\phi} \dot{\phi}
+ V^{\prime}(\phi) = 0~,
\label{eq:evolution}
\end{equation}
where the prime denotes derivation with respect to $\phi$
and $\Gamma_{\phi}$ is the decay width \cite{width} of
the inflaton. Assume, for the moment, that the decay time
of $\phi $, $t_d = \Gamma_{\phi}^{-1}$, is much greater
than $H^{-1}$, the expansion time for inflation.
Then the term $\Gamma_{\phi} \dot{\phi}$ can be ignored
and Eq.(\ref{eq:evolution}) reduces to
\begin{equation}
\ddot{\phi} + 3 H \dot{\phi} + V^{\prime}(\phi) = 0~.
\label{eq:reduce}
\end{equation}
Inflation is by definition the situation where
$\ddot{\phi}$ is subdominant to the `friction' term
$3H \dot{\phi}$ in this equation (and the kinetic energy
density is subdominant to the potential one).
Eq.(\ref{eq:reduce}) then further reduces to the inflationary
equation \cite{slowroll}
\begin{equation}
3H \dot{\phi} = - V^{\prime} (\phi)~,
\label{eq:infeq}
\end{equation}
which gives
\begin{equation}
\ddot{\phi} =
- \frac {V^{\prime\prime}(\phi)\dot{\phi}}
{3H(\phi)} + \frac {V^{\prime}(\phi)}
{3H^{2}(\phi)} H^\prime (\phi) \dot{\phi}~.
\label{eq:phidd}
\end{equation}
Comparing the two terms in the rhs of this equation with
the `friction' term in Eq.(\ref{eq:reduce}), we get the
conditions for inflation (slow roll conditions):
\begin{equation}
\eta \equiv \frac{M_{P}^{2}}{8 \pi} \bigg |
\frac {V^{\prime\prime}(\phi)}
{V(\phi)} \bigg | \leq 1~,
~\epsilon \equiv \frac {M_{P}^{2}}{16 \pi}
\left(\frac {V^{\prime}(\phi)}
{V(\phi)}\right)^{2} \leq 1~.
\label{eq:src}
\end{equation}
The end of the slow roll over occurs when either of the
these inequalities is saturated. If
$\phi_f$ is the value of $\phi$ at the end of inflation,
then $t_f \sim H^{-1}(\phi_f)$.

\par
The number of e-foldings during inflation can be calculated
as follows:
$$
N(\phi_{i}\rightarrow \phi_{f}) \equiv \ell n
\left(\frac {a(t_{f})}
{a(t_{i})}\right) = \int^{t_{f}} _{t_{i}} Hdt=
$$
\begin{equation}
 \int^{\phi_{f}}_{\phi_{i}}
\frac {H (\phi)}{\dot{\phi}} d \phi = -
\int^{\phi_{f}}_{\phi_{i}}
\frac {3 H^2 (\phi) d \phi}
{V^{\prime}(\phi)}~,
\label{eq:nefolds}
\end{equation}
where  Eqs.(\ref{eq:efold}), (\ref{eq:infeq}) and the
definition of $H = \dot{a}/a$ were used. For simplicity,
we can shift the field $\phi$ so that the global minimum of
the potential is displaced at $\phi$ = 0. Then, if
$V(\phi) = \lambda \phi^{\nu}$ during inflation, we have
$$
N(\phi_{i} \rightarrow \phi_{f}) =
- \int^{\phi_{f}}_{\phi_{i}} \frac
{3H^2(\phi)d\phi}{V^{\prime}(\phi)} =
$$
\begin{equation}
- 8 \pi G \int^{\phi_{f}}_{\phi_{i}} \frac
{V(\phi)d\phi}{V^{\prime}(\phi)}=\frac {4 \pi G}{\nu}
(\phi^{2}_{i}-\phi^{2}_{f})~.
\label{eq:expefold}
\end{equation}
Assuming that $\phi_{i} \gg \phi_{f}$, this reduces to
$N(\phi) = (4 \pi G/\nu)\phi^2$.

\section{Coherent Field Oscillations}
\label{sec:osci}

After the end of inflation at cosmic time $t_f$, the term
$\ddot{\phi}$ takes over and Eq.(\ref{eq:reduce}) reduces
to $\ddot{\phi} + V^{\prime}(\phi)=0$, which means that
$\phi$ starts oscillating coherently about the global
minimum of the potential. In reality, due to the `friction'
term, $\phi$ performs damped oscillations with a rate of
energy density loss given by
\begin{equation}
\dot{\rho}= \frac {d}{dt}\left(\frac{1}{2}
\dot{\phi}^2 + V(\phi)\right)
= - 3H \dot{\phi}^2=-3H(\rho+p)~,
\label{eq:damp}
\end{equation}
where $\rho = \dot{\phi}^2/2 + V(\phi) $ and the
pressure $p=\dot{\phi}^2/2 - V(\phi)$. Averaging $p$
over one oscillation of $\phi$ and writing
\cite{oscillation} $\rho+p=\gamma \rho$, we get
$\rho \propto a^{-3 \gamma}$ and $a(t) \propto
t^{2/3 \gamma}$ (see Sec.\ref{subsec:friedmann}).

\par
The number $\gamma$ for an oscillating
field can be written as (assuming a symmetric potential)
\begin{equation}
\gamma = \frac {\int^{T}_{0} \dot{\phi}^{2} dt}
{\int^{T}_{0} \rho dt} =
\frac{\int^{\phi_{{\rm{max}}}}_{0}
\dot{\phi}d \phi}
{\int^{\phi_{{\rm{max}}}}_{0}
(\rho/\dot{\phi}) d\phi}~,
\label{eq:gamma}
\end{equation}
where $T$ and $\phi_{{\rm{max}}}$ are the period and
the amplitude of the oscillation respectively. From the
equation $\rho = \dot{\phi}^2/2 + V(\phi)=
V_{{\rm{max}}}$, where $V_{{\rm{max}}}$ is the
maximal potential energy density, we obtain $\dot{\phi}=
\sqrt{2(V_{{\rm{max}}} - V(\phi))}$.
Substituting this in Eq.(\ref{eq:gamma}) we get
\cite{oscillation}
\begin{equation}
\gamma = \frac {2 \int^{\phi_{{\rm{max}}}}_{0}
(1-V/V_{{\rm{max}}})^{1/2}
d\phi}{\int^{\phi_{{\rm{max}}}}_{0}
(1-V/V_{{\rm{max}}})^{-1/2} d \phi}~~\cdot
\label{eq:gammafinal}
\end{equation}
For a potential of the simple form $V(\phi) =
\lambda \phi^{\nu}$,~$\gamma$ is
readily found to be given by $\gamma=2 \nu/(\nu +2)$.
Consequently, in this case, $\rho \propto
a^{-6\nu/(\nu+2)}$ and $a(t) \propto t^{(\nu+2)/3 \nu}$.
For $\nu=2$, in particular, one has $\gamma$=1,
~$\rho \propto a^{-3}$,~$a(t)\propto t^{2/3}$ and
the oscillating field behaves like pressureless `matter'.
This is not unexpected since a coherent oscillating massive
free field corresponds to a distribution of static massive
particles. For $\nu$=4, however, we obtain $\gamma = 4/3$,
~$\rho \propto a^{-4}$,~$a(t) \propto t^{1/2}$ and the
system resembles `radiation'. For $\nu = 6$, one has
$\gamma=3/2$,~$ \rho \propto a^{-4.5}$,~$a(t)
\propto t^{4/9}$ and the expansion is slower than in a
radiation dominated universe (the pressure is higher than
in `radiation').

\section{Decay of the Field $\phi$}
\label{sec:decay}

Reintroducing the `decay' term
$\Gamma_{\phi} \dot{\phi}$,
Eq.(\ref{eq:evolution}) can be written as
\begin{equation}
\dot{\rho} =
\frac{d}{dt} \left(\frac{1}{2} \dot{\phi}^2 +
V(\phi)\right) = - (3H + \Gamma_\phi)\dot{\phi}^2~,
\label{eq:decay}
\end{equation}
which is solved \cite{reheat,oscillation} by
\begin{equation}
\rho(t)=\rho_{f}
\left(\frac{a(t)}{a(t_{f})}\right)^{-3 \gamma}
{\rm{exp}} [ -\gamma \Gamma_{\phi}(t-t_f)]~,
\label{eq:rho}
\end{equation}
where $\rho_f$ is the energy density at the end of
inflation at cosmic time $t_f$. The second and third
factors in the rhs of this equation represent the dilution
of the field energy due to the expansion of the universe
and the decay of $\phi$ to light particles respectively.

\par
All pre-existing `radiation' (known as `old radiation')
was diluted by inflation, so the only `radiation' present
is the one produced by the decay of $\phi$ and is known as
`new radiation'. Its energy density satisfies
\cite{reheat,oscillation} the equation
\begin{equation}
\dot{\rho}_{r} = - 4 H \rho_{r} +
\gamma \Gamma_{\phi} \rho~,
\label{eq:newrad}
\end{equation}
where the first term in the rhs represents the dilution of
radiation due to the cosmological expansion while the second
one is the energy density transfer from $\phi$ to
`radiation'. Taking $\rho_{r}(t_f)$=0, this equation gives
\cite{reheat,oscillation}
$$
\rho_{r}(t) = \rho_{f}\left(\frac {a(t)}
{a(t_{f})}\right)^{-4}
$$
\begin{equation}
\int^{t}_{t_{f}}
\left(\frac{a(t^{\prime})}
{a(t_{f})}\right)^{4-3 \gamma}
e^{ -\gamma \Gamma_{\phi} (t^{\prime}-t_f)}
~\gamma \Gamma_{\phi} dt^{\prime}~.
\label{eq:rad}
\end{equation}
For $t_{f} \ll t_{d}$ and $\nu =2$, this expression is
approximated by
\begin{equation}
\rho_{r}(t)=\rho_{f}\left(\frac {t}{t_f}\right)^{-8/3}
\int^{t}_{0}
\left(\frac{t^{\prime}}{t_{f}}\right)^{2/3}
e^{-\Gamma_{\phi}t^{\prime}} dt^{\prime}~,
\label{eq:appr}
\end{equation}
which, using the formula
\begin{equation}
\int_{0}^{u} x^{p-1} e^{-x}dx =
e^{-u}~\sum^{\infty}_{k=0}~
\frac {u^{p+k}}{p(p+1)\cdot\cdot\cdot(p+k)}~~,
\label{eq:formula}
\end{equation}
can be written as
\begin{equation}
\rho_{r} = \frac {3}{5}~\rho~\Gamma_{\phi}t
\left[1 + \frac {3}{8}~\Gamma_{\phi}t
+ \frac {9}{88}~(\Gamma_{\phi}t)^2+ \cdots \right]~,
\label{eq:expand}
\end{equation}
with $\rho = \rho_{f} (t/t_{f})^{-2}{\rm{exp}}
(-\Gamma_{\phi}t)$ being the energy density of the field
$\phi$ which performs damped oscillations and decays into
`light' particles.

\par
The energy density of the `new radiation' grows relative
to the energy density of the oscillating field and becomes
essentially equal to it at a cosmic time $t_{d} =
\Gamma_{\phi}^{-1}$ as one can deduce from
Eq.(\ref{eq:expand}). After this time, the universe enters
into the radiation dominated era and the normal big bang
cosmology is recovered. The temperature at $t_{d},
~T_{r}(t_{d})$, is historically called the `reheat'
temperature although no supercooling and subsequent
reheating of the universe actually takes place. Using the
time to temperature relation in Eq.(\ref{eq:temptime}) for
a radiation dominated universe we find that
\begin{equation}
T_{r} = \left(\frac {45}{16 \pi^{3}g_*}\right)^{1/4}
(\Gamma_{\phi} M_{P})^{1/2}~,
\label{eq:reheat}
\end{equation}
where $g_*$ is the effective number of degrees of freedom.
For a potential of the type $V(\phi)=\lambda\phi^{\nu}$,
the total expansion of the universe during the period of
damped field oscillations is
\begin{equation}
\frac{a(t_{d})}{a(t_{f})} = \left(
\frac{t_{d}}{t_{f}}\right)^{\frac{\nu + 2}{3\nu}}~.
\label{eq:expansion}
\end{equation}

\section{Density Perturbations}
\label{sec:density}

We are ready to sketch how inflation solves the density
fluctuation problem described in Sec.\ref{subsec:fluct}.
As a matter of fact, inflation not only homogenizes the
universe but also provides us with the primordial density
fluctuations necessary for the structure formation in the
universe. To understand the origin of these fluctuations,
we must first introduce the notion of `event horizon'. Our
`event horizon', at a cosmic time $t$, includes all points
with which we will eventually communicate sending signals
at $t$. The `instantaneous' (at cosmic time $t$) radius of
the `event horizon' is
\begin{equation}
d_{e}(t) = a(t) \int ^{\infty}_{t}
\frac{dt^{\prime}}{a(t^{\prime})}~\cdot
\label{eq:event}
\end{equation}
It is obvious, from this formula, that the `event horizon'
is infinite for matter or radiation dominated universe. For
inflation, however, we obtain a slowly varying `event horizon'
with radius $d_{e}(t) = H^{-1} < \infty$. Points, in our
`event horizon' at $t$, with which we can communicate sending
signals at $t$, are eventually pulled away by the
`exponential' expansion and we cease to be able to communicate
with them again emitting signals at later times. We say that
these points (and the corresponding scales) crossed outside
the `event horizon'. The situation is very similar to that of
a black hole. Indeed, the exponentially expanding (de Sitter)
space is like a black hole turned inside out. This means that
we are inside and the black hole surrounds us from all sides.
Then, exactly as in a black hole, there are quantum
fluctuations of the `thermal type' governed by the
`Hawking temperature' \cite{hawking,gibbons}
$T_{H} = H/2\pi$. It turns out \cite{bunch,vilenkin} that
the quantum fluctuations of all massless fields (the
inflaton is nearly massless due to the `flatness' of the
potential) are $\delta \phi = H/ 2\pi = T_{H}$. These
fluctuations of $\phi$ lead to energy density fluctuations
$\delta \rho = V^{\prime} (\phi) \delta \phi$. As the
scale of this perturbations crosses outside the `event
horizon', they become \cite{fischler} classical metric
perturbations.

\par
The evolution of these fluctuations outside the
`inflationary horizon' is quite subtle and involved due
to the gauge freedom in general relativity. However, there
is a simple gauge invariant quantity \cite{zeta} $\zeta
\approx \delta \rho/(\rho +p)$, which remains constant
outside the horizon. Thus, the density fluctuation at any
present physical (`comoving') scale $\ell,
~(\delta \rho/\rho)_{\ell}$, when this scale crosses
inside the post-inflationary particle horizon ($p$=0 at
this instance) can be related to the value of $\zeta$ when
the same scale crossed outside the inflationary `event
horizon' (symbolically at $\ell \sim H^{-1}$). This latter
value of $\zeta$ can be found using Eq.(\ref{eq:infeq}) and
turns out to be
$$
\zeta \mid_{\ell \sim H^{-1}} =
\left(\frac {\delta \rho}{\dot {\phi}^2}\right)
_{\ell \sim H^{-1}} =
\left(\frac {V^{\prime} (\phi) H(\phi)}
{2 \pi \dot{\phi}^2}\right)_{\ell \sim H^{-1}}
$$
\begin{equation}
=- \left(\frac {9 H^{3}(\phi)}
{2 \pi V^{\prime}(\phi)}\right)_{\ell \sim H{-1}}~\cdot
\label{eq:zeta}
\end{equation}
Taking into account an extra 2/5 factor from the fact that
the universe is matter dominated when the scale $\ell$
re-enters the horizon, we obtain
\begin{equation}
\left(\frac {\delta \rho}{\rho}\right)_{\ell} =
\frac {16 \sqrt {6 \pi}}{5}~\frac
{V^{3/2}(\phi_{\ell})}{M^{3}_{P} V^{\prime}
(\phi_{\ell})} ~\cdot
\label{eq:deltarho}
\end{equation}

\par
The calculation of $\phi_{\ell}$, the value of the
inflaton field when the `comoving' scale $\ell$ crossed
outside the `event horizon', goes as follows. A `comoving'
(present physical) scale $\ell$, at $T_r$, was equal to
$\ell (a(t_{d})/a(t_{0})) = \ell(T_{0}/T_{r})$.
Its magnitude at the end of inflation ($t=t_{f}$) was
equal to $\ell (T_{0}/T_{r})(a(t_{f})/a(t_{d}))
= \ell (T_{0}/T_{r})(t_{f}/t_{d})^{(\nu +2)/3 \nu}$
$\equiv \ell_{{\rm{phys}}}(t_{f})$, where the potential
$V(\phi)=\lambda\phi^{\nu}$ was assumed. The scale
$\ell$, when it crossed outside the inflationary horizon,
was equal to $H^{-1}(\phi_{\ell})$. We, thus, obtain
\begin{equation}
H^{-1}(\phi_{\ell}) e^{N(\phi_{\ell})} =
\ell_{{\rm{phys}}}(t_{f})~.
\label{eq:lphys}
\end{equation}
Solving this equation, one can calculate $\phi_{\ell}$
and, thus, $N(\phi_{\ell})\equiv N_{\ell}$, the number
of e-foldings the scale $\ell$ suffered during inflation.
In particular, for our present horizon scale
$\ell \approx 2H_{0}^{-1} \sim10^4$~Mpc, it turns out
that $N_{H_{0}}\approx 50-60$.

\par
Taking the potential $V(\phi)= \lambda \phi^4$,
~Eqs.(\ref{eq:expefold}), (\ref{eq:deltarho}) and
(\ref{eq:lphys}) give
$$
\left(\frac {\delta \rho} {\rho}\right)_{\ell} =
\frac {4 \sqrt{6 \pi}} {5}\lambda^{1/2}
\left(\frac{\phi_{\ell}}{M_{P}}\right)^3 =
$$
\begin{equation}
\frac {4 \sqrt{6 \pi}}{5}\lambda^{1/2}
\left(\frac {N_{\ell}}{\pi}\right)^{3/2}~\cdot
\label{eq:nl}
\end{equation}
The measurements of COBE~\cite{cobe}, $ (\delta \rho/
\rho)_{H_{0}} \approx 6 \times 10^{-5}$, then imply
that $\lambda \approx 6 \times 10^{-14}$ for
$N_{H_{0}} \approx 55$. Thus, we see that the inflaton
must be a very weakly coupled field. In nonsupersymmetric
GUTs, the inflaton is necessarily gauge singlet since
otherwise radiative corrections will certainly make
it strongly coupled. This is, undoubtedly, not a very
satisfactory situation since we are forced to introduce
an otherwise unmotivated extra {\it ad hoc} very weakly
coupled gauge singlet. In supersymmetric GUTs, however,
the inflaton could be identified \cite{nonsinglet} with
a conjugate pair of gauge nonsinglet fields $\phi$,
$\bar{\phi}$, already existing in the theory and causing
the gauge symmetry breaking. Absence of strong radiative
corrections from gauge interactions is guaranteed, in this
case, by the mutual cancellation of the D terms of these
fields.

\par
The spectrum of density fluctuations which emerge from
inflation can also be analyzed. We will again take the
potential $V(\phi) =\lambda \phi^{\nu}$. One then finds
that $(\delta \rho/ \rho)_{\ell}$ is proportional to
$\phi_{\ell}^{(\nu+2)/2}$ which, combined with the fact
that $N(\phi_{\ell})$ is proportional to
$\phi_{\ell}^{2}$ (see Eq.(\ref{eq:expefold})), gives
\begin{equation}
\left(\frac {\delta \rho}{\rho}\right)_{\ell} =
\left(\frac{\delta \rho}{\rho}\right)_{H_{0}}\left(
\frac{N_{\ell}}{N_{H_{0}}}\right)^{\frac{\nu+2}{4}}.
\label{eq:spectrum}
\end{equation}
The scale $\ell$ divided by the size of our present horizon
($\approx 10^4~{\rm{Mpc}}$) should equal exp$(N_{\ell}-
N_{H_{0}})$. This gives $N_{\ell}/N_{H_{0}} =
1 + \ell n(\ell/10^{4})^{1/N_{H_{0}}}$ which expanded
around $\ell \approx 10^4$~Mpc and substituted in
Eq.(\ref{eq:spectrum}) yields
\begin{equation}
\left(\frac{\delta \rho}{\rho}\right)_{\ell} =
\left(\frac{\delta \rho}{\rho}\right)_{H_{0}}\left(
\frac {\ell}{10^4~{\rm{Mpc}}}\right)^{\alpha_{s}}~,
\label{eq:alphas}
\end{equation}
with $\alpha_{s}=(\nu + 2) /4 N_{H_{0}}$. For $\nu=4$,
$\alpha_{s} \approx 0.03$ and, thus,  the density
fluctuations are essentially scale independent.

\section{Density Fluctuations in `Matter'}
\label{sec:matter}

We will now discuss the evolution of the primordial density
fluctuations after their scale enters the post-inflationary
horizon. To this end, we introduce \cite{bardeen} the
`conformal' time, $\eta$, so that the Robertson-Walker
metric takes the form of a conformally expanding Minkowski
space:
\begin{equation}
ds^2 = -dt^2 + a^2 (t)~d\bar{r}^2 = a^{2}(\eta)~
(-d \eta^2 + d\bar{r}^2)~,
\label{eq:conf}
\end{equation}
where $\bar{r}$ is a `comoving' 3-vector. The Hubble
parameter now takes the form $H\equiv \dot{a}(t)/a(t)
=a^{\prime}(\eta)/a^{2}(\eta)$ and the Friedmann
Eq.(\ref{eq:friedmann}) can be rewritten as
\begin{equation}
\frac{1}{a^{2}}\left(\frac{a^{\prime}}{a}\right)^{2}
=\frac {8 \pi G}{3} \rho~,
\label{eq:conffried}
\end{equation}
where primes denote derivation with respect to the
`conformal' time $\eta$. The continuity
Eq.(\ref{eq:continuity}) takes the form $ \rho^{\prime}
= - 3 \tilde {H} (\rho+p)$ with $\tilde{H} =
a^{\prime}/a$. For a matter dominated universe, $\rho
\propto a^{-3}$ which gives $a=(\eta/\eta_{0})^2$
and $a^{\prime}/a =2/\eta$ ($\eta_0$ is the present
value of $\eta$).

\par
The Newtonian Eq.(\ref{eq:newton}) can now be written in
the form
\begin{equation}
\delta^{\prime \prime}_{\bar{k}} (\eta) +
\frac {a^{\prime}}{a}
\delta^{\prime}_{\bar{k}}(\eta)
- 4 \pi G \rho a^{2} \delta_{\bar{k}} (\eta) =0~,
\label{eq:confnewton}
\end{equation}
and the growing (Jeans unstable) mode
$\delta_{\bar{k}} (\eta)$
is proportional to $\eta^{2}$ and can be expressed
\cite{schaefer} as
\begin{equation}
\delta_{\bar{k}} (\eta) = \epsilon_{H} \left(
\frac {k \eta}{2}\right)^{2} \hat {s} (\bar{k})~,
\label{eq:growmode}
\end{equation}
where $\hat{s}(\bar{k})$ is a Gaussian random variable
satisfying
\begin{equation}
<\hat{s}(\bar {k})> = 0~,~<\hat{s}(\bar {k})
\hat {s}(\bar {k}^{\prime})>
= \frac {1} {k^{3}} \delta (\bar{k} -
\bar {k}^{\prime})~,
\label{eq:gauss}
\end{equation}
and $\epsilon_{H}$ is the amplitude of the perturbation
when its scale crosses inside the post-inflationary horizon.
The latter can be seen as follows. A `comoving' (present
physical) length $\ell$ crosses inside the post-inflationary
horizon when  $a \ell/ 2 \pi = H^{-1}=a^2/a^{\prime}$
which gives $\ell/ 2\pi \equiv k^{-1} = a/a^{\prime}=
\eta_{H}/2$ or $k \eta_{H}/2 = 1$, where $\eta_{H}$ is
the `conformal' time at horizon crossing. This means that, at
horizon crossing, $\delta_{\bar{k}} (\eta_{H}) =
\epsilon_{H}\hat{s} (\bar{k})$. For scale invariant
perturbations, the amplitude $\epsilon _{H}$ is constant.
The gauge invariant perturbations of the scalar gravitational
potential are given \cite{bardeen} by the Poisson's equation,
\begin{equation}
\Phi = - 4 \pi G \frac {a^2}{k^2}\rho
\delta_{\bar{k}}(\eta)~.
\label{eq:poisson}
\end{equation}
From the  Friedmann Eq.(\ref{eq:conffried}), we then obtain
\begin{equation}
\Phi = - \frac {3}{2} \epsilon_{H} \hat{s} (\bar{k})~.
\label{eq:scalarpot}
\end{equation}

\par
The spectrum of the density perturbations can be
characterized by the correlation function ($\bar{x}$ is
a `comoving' 3-vector)
\begin{equation}
\xi(\bar{r}) \equiv
<\tilde{\delta}^{*}(\bar{x},\eta)
\tilde{\delta}(\bar{x} + \bar{r},
\eta)>~,
\label{eq:corr}
\end{equation}
where
\begin{equation}
\tilde{\delta}(\bar{x},\eta) = \int d^{3}k
\delta_{\bar{k}}(\eta)e^{i\bar{k}\bar{x}}~.
\label{eq:fourier}
\end{equation}
Substituting Eq.(\ref{eq:growmode}) in Eq.(\ref{eq:corr})
and then using Eq.(\ref{eq:gauss}), we obtain
\begin{equation}
\xi (\bar{r}) = \int d^{3}k e^{-i\bar{k}\bar{r}}
\epsilon^2_{H}
\left(\frac {k \eta}{2}\right)^{4} \frac{1}{k^{3}}~,
\label{eq:index}
\end{equation}
and the spectral function  $P(k, \eta) =
\epsilon^2_{H}(\eta^{4}/16)k$ is proportional to $k$
for $\epsilon_{H}$ constant. We say that, in this case,
the `spectral index' $n=1$ and we have a Harrison-Zeldovich
\cite{hz} flat spectrum. In the general case,
$P \propto k^{n}$ with $n = 1-2 \alpha_{s}$ (see
Eq.(\ref{eq:alphas})). For $V(\phi)= \lambda \phi^{4}$,
we get $n \approx 0.94$.

\section{Temperature Fluctuations}
\label{sec:temperature}

The density inhomogeneities produce temperature fluctuations
in the CBR. For angles
$\theta \stackrel{_{>}}{_{\sim }} 2~^o$, the dominant
effect is the scalar Sachs-Wolfe \cite{sachswolfe} effect.
Density perturbations on the `last scattering surface'
cause scalar gravitational potential fluctuations, $\Phi$,
which, in turn, produce temperature fluctuations in the CBR.
The physical reason is that regions with a deep gravitational
potential will cause the photons to lose energy as they climb
up the well and, thus, appear cooler. For
$\theta \stackrel{_{<}}{_{\sim }} 2~^o$, the dominant
effects are: i) Motion of the last scattering surface causing
Doppler shifts, and ii) Intrinsic fluctuations of the photon
temperature, $T_{\gamma}$, which are more difficult to
calculate since they depend on microphysics, the ionization
history, photon streaming and other effects.

\par
The temperature fluctuations at an angle $\theta$ due to
the scalar Sachs-Wolfe effect turn out \cite{sachswolfe}
to be  $(\delta T/T)_{\theta} = - \Phi_{\ell}/3$,
$\ell$ being the `comoving' scale on the `last scattering
surface' which subtends the angle $\theta$
[~$\ell \approx 100~h^{-1}(\theta/{\rm{degrees}})
~{\rm{Mpc}}~]$ and $\Phi_{\ell}$ the corresponding
scalar gravitational potential fluctuations. From
Eq.(\ref{eq:scalarpot}), we then obtain
$(\delta T/T)_{\theta} =
(\epsilon_{H}/2) \hat{s} (\bar{k})$, which using
Eq.(\ref{eq:growmode}) gives the relation
\begin{equation}
\left(\frac{\delta T}{T}\right)_{\theta} =
\frac{1}{2} \delta_{\bar{k}}(\eta_{H})
= \frac {1}{2} \left(\frac{\delta \rho}
{\rho}\right)_{\ell \sim 2\pi k^{-1}}\cdot
\label{eq:swe}
\end{equation}
The COBE scale (present horizon) corresponds to $\theta
\approx 60~^o$. Eqs.(\ref{eq:expefold}),
(\ref{eq:deltarho}) and (\ref{eq:swe}) give
\begin{equation}
\left(\frac {\delta T} {T}\right)_{\ell}
\propto \left(\frac{\delta \rho}{\rho}\right)_{\ell}
\propto
\frac{V^{3/2}(\phi_{\ell})}{M^{3}_{P} V^{\prime}
(\phi_{\ell})} \propto N_{\ell}^{\frac{\nu+2}{4}}~.
\label{ewq:tempflu}
\end{equation}
Analyzing the temperature fluctuations in spherical harmonics,
the quadrupole anisotropy due to the scalar Sachs-Wolfe effect
can be obtained:
\begin{equation}
\left(\frac{\delta T}{T}\right)_{Q-S} =
\left(\frac{32 \pi}
{45}\right)^{1/2}
\frac{V^{3/2}(\phi_{\ell})}{M^{3}_{P}
V^{\prime}(\phi_{\ell})}~\cdot
\label{eq:quadrupole}
\end{equation}
For $V(\phi) = \lambda \phi^{\nu}$, this becomes
$$
\left(\frac{\delta T}{T}\right)_{Q-S} =
\left(\frac{32 \pi}{45}\right)^{1/2}
\frac{\lambda^{1/2}\phi_{\ell}^{\frac{\nu+2}{2}}}
{\nu M^{3}_{P}}=
$$
\begin{equation}
\left(\frac{32 \pi}{45}\right)^{1/2}
\frac {\lambda^{1/2}}{\nu M^{3}_{P}}
\left(\frac{\nu M^{2}_{P}}
{4 \pi}\right)^{\frac{\nu+2}{4}}
N_{\ell}^{\frac{\nu+2}{4}}~.
\label{eq:anisotropy}
\end{equation}
Comparing this with COBE \cite{cobe} measurements,
$(\delta T/T)_{Q} \approx 6.6 \times 10^{-6}$, we
obtain $\lambda \approx 6 \times 10^{-14}$, for $\nu=4$,
and number of e-foldings suffered by our present horizon
scale during the inflationary phase
$N_{\ell \sim H^{-1}_{0}} \equiv N_{Q} \approx 55$.

\par
There are also `tensor' \cite{tensor} fluctuations in
the temperature of CBR. The quadrupole tensor anisotropy
is
\begin{equation}
\left(\frac{\delta T}{T}\right)_{Q-T}
\approx 0.77~\frac
{V^{1/2}(\phi_{\ell})}{M^{2}_{P}}~\cdot
\label{eq:tensor}
\end{equation}
The total quadrupole anisotropy is given by
\begin{equation}
\left(\frac{\delta T}{T}\right)_{Q} =
\left[ \left(\frac{\delta T}{T}\right)
^{2}_{Q-S} + \left(\frac{\delta T}{T}\right)^{2}
_{Q-T}\right]^{1/2},
\label{eq:total}
\end{equation}
and the ratio
\begin{equation}
r = \frac{\left(\delta T/T\right)^{2}_{Q-T}}
{\left(\delta T/T\right)^{2}_{Q-S}} \approx 0.27
~\left(\frac{M_{P} V^{\prime}(\phi_{\ell})}
{V(\phi_{\ell})}\right)^{2}\cdot
\label{eq:ratio}
\end{equation}
For $V(\phi) = \lambda \phi^{\nu}$, we obtain $r
\approx 3.4~\nu/N_{H}\ll 1$, and the `tensor'
contribution to the temperature fluctuations of the CBR
is negligible.

\section{Hybrid Inflation}
\label{sec:hybrid}

\subsection{The non Supersymmetric Version}
\label{subsec:nonsusy}

The most important disadvantage of the inflationary
scenarios described so far is that they need extremely
small coupling constants in order to reproduce the results
of COBE \cite{cobe}. This difficulty was overcome some
years ago by Linde \cite{hybrid} who proposed, in the
context of nonsupersymmetric GUTs, an inflationary scenario
known as hybrid inflation. The  idea was to use two real
scalar fields $\chi$ and $\sigma$ instead of one that was
normally used. The  field $\chi$ provides the vacuum energy
which drives inflation  while  $\sigma$ is the slowly
varying field during inflation. The main advantage of this
scenario is that it can reproduce the observed temperature
fluctuations of the CBR with `natural' values of the
parameters in contrast to previous realizations of inflation
(like the new \cite{new} or chaotic \cite{chaotic}
inflationary scenarios).

\par
The potential utilized by Linde is
\begin{equation}
V ( \chi, \sigma)= \kappa^2 \left( M^2 -
\frac {\chi^2}{4}\right)^2 +   \frac{\lambda^2
\chi^2 \sigma^2}   {4}   +   \frac {m^2\sigma^2}{2}~~,
\label{eq:lindepot}
\end{equation}
where  $\kappa,~\lambda$ are dimensionless positive
coupling constants and $M$, $m$ are mass parameters.
The vacua lie  at $\langle \chi\rangle= \pm 2 M$,
$\langle \sigma \rangle=0$. Putting
$m$=0, for  the moment, we observe that the potential
possesses an  exactly flat direction  at  $\chi=0$
with $V(\chi=0 ,\sigma)=\kappa^2 M^4$. The mass
squared of the field $\chi$ along this flat direction
is given by $m^2_\chi=- \kappa^2 M^2 +
\lambda^2  \sigma^2/2$  and  remains nonnegative for
$\sigma \geq \sigma_c = \sqrt  {2}  \kappa M/
\lambda $. This means that, at $\chi=0$ and
$\sigma \geq  \sigma_c$, we obtain a valley of minima with
flat bottom. Reintroducing  the mass parameter $m$ in
Eq.(\ref{eq:lindepot}), we observe that this valley acquires a
nonzero slope. A region of the universe, where $\chi$ and
$\sigma$  happen to be almost uniform with negligible kinetic
energies and  with  values close to the bottom of the valley
of minima, follows  this valley in its subsequent  evolution
and undergoes  inflation.

\par
The quadrupole anisotropy of CBR
produced during this hybrid inflation can be estimated, using
Eq.(\ref{eq:quadrupole}), to be
\begin{equation}
\left(\frac {\delta T}{T}\right)_{Q}
\approx  \left(\frac {16 \pi}{45}\right)^{1/2}
\frac{\lambda \kappa^2 M^5}
{M^3_Pm^2}~.
\label{eq:lindetemp}
\end{equation}
The COBE~\cite{cobe} result,
$(\delta T/T)_{Q} \approx 6.6 \times 10^{-6}$, can then
be reproduced with $M \approx 2.86 \times 10^{16}$ GeV (the supersymmetric GUT vev) and $m \approx 1.3~\kappa
\sqrt {\lambda}\times 10^{15}$ GeV $ \sim 10^{12}$ GeV
for $\kappa, \lambda \sim 10^{-2}$.

\par
Inflation terminates abruptly
at $\sigma=\sigma_{c}$ and is followed by a `waterfall',
i.e., a sudden entrance into  an oscillatory phase about a
global minimum. Since the system can fall into either of the
two available global minima with equal probability,
topological defects are copiously produced if they are
predicted by the particular particle physics model one is
considering.

\subsection{The Supersymmetric Version}
\label{subsec:susy}

The hybrid inflationary scenario is \cite{lyth}
`tailor made' for application to supersymmetric GUTs except
that the mass of $\sigma$, $m$, is unacceptably large for
supersymmetry, where all scalar fields acquire masses of
order $m_{3/2} \sim 1$ TeV (the gravitino mass) from
soft supersymmetry breaking.

\par
To see this, consider a supersymmetric GUT with a
(semi-simple) gauge group $G$ of rank $\geq 5$ with
$G \to G_S$ (the standard model gauge group) at a scale
$M \sim 10^{16}~$GeV. The spectrum of the theory below
$M$ is assumed to coincide with the spectrum of the
minimal supersymmetric
standard model (MSSM) plus standard model singlets
so that the successful predictions for $\alpha_{s}$,
${\rm{sin}}^{2} \theta_{W}$ are retained. The theory
may also possess global symmetries. The breaking of $G$
is achieved through the superpotential
\begin {equation}
W = \kappa S (- M^2 + \bar{\phi} \phi),
\label{eq:superpot}
\end {equation}
where $\bar{\phi}, \phi$ is a conjugate pair of $G_{S}$
singlet left handed superfields belonging to
nontrivial representations of $G$ and reduce its rank by
their vevs and $S$ is a gauge singlet left handed superfield.
The coupling constant $\kappa$ and the mass parameter $M$
can be made positive by phase redefinitions. This
superpotential has the most general form consistent with
a $U(1)$ R-symmetry under which $W \to e^{i\theta} W,
~S \to e^{i \theta}S,~\bar {\phi} \phi \to
\bar{\phi} \phi$.

\par
The potential derived from the superpotential
$W$ in Eq.(\ref{eq:superpot}) is
\begin{eqnarray*}
V=\kappa^2 \mid M^2 - \bar{\phi} \phi \mid^2
 +\kappa^2 \mid S \mid^2
(\mid \phi \mid^2  + \mid \bar{\phi}\mid^2)
\end{eqnarray*}
\begin{equation}
+{\rm{ D-terms}}.
\label{eq:hybpot}
\end{equation}
Restricting ourselves to the D flat direction
$\bar{\phi}^* =\phi$ which contains the supersymmetric
vacua and performing appropriate gauge and R-
transformations, we can bring $S$, $\bar{\phi}$, $\phi$
on the real axis, i.e., $S \equiv \sigma/\sqrt{2}$,
$\bar{\phi}=\phi \equiv \chi/2$, where $\sigma$,
$\chi$ are normalized real scalar fields. The potential
then takes the form in Eq.(\ref{eq:lindepot}) with
$\kappa = \lambda$ and $m=0$ and, thus, Linde's
potential for hybrid inflation is almost obtainable
from supersymmetric GUTs but without the mass term
of $\sigma$ which is, however, of crucial importance
since it provides the slope of the valley of minima
necessary for driving the inflaton towards the vacua.

\par
One way to obtain a valley of minima useful for inflation
is \cite{lp} to replace the renormalizable trilinear
superpotential term in Eq.(\ref{eq:superpot}) by the
next order nonrenormalizable coupling. Another way, which
we will adopt here, is \cite{dss} to keep the
renormalizable superpotential in Eq.(\ref{eq:superpot}) and
use the radiative corrections along the inflationary valley
($\bar{\phi}=\phi = 0$~, $S > S_{c} \equiv M$). In fact,
the breaking of supersymmetry by the `vacuum' energy density
$\kappa^{2} M^{4}$ along this valley causes a mass
splitting in the supermultiplets $\bar{\phi}$, $\phi$.
This results to the existence of important radiative
corrections on the inflationary valley. At one-loop,
and for $S$ sufficiently larger than $S_{c}$, the
inflationary potential is given
\cite{dss,lss} by
$$
V_{{\rm{eff}}} (S) = \kappa^2 M^4
$$
\begin{equation}
\left[ 1 + \frac{\kappa^2}{16\pi^2} \left( \ln
\left(\frac{\kappa^{2} S^{2}}{\Lambda^2}\right)
+ \frac{3}{2} - \frac{S_c^4}{12S^4} +
\cdots \right)\right]~,
\label{eq:veff}
\end{equation}
where $\Lambda$ is a suitable mass renormalization scale.

\par
From Eqs.(\ref{eq:veff}) and (\ref{eq:quadrupole}),
we find the cosmic microwave quadrupole anisotropy:
\begin{equation}
\left(\frac{\delta T}{T}\right)_{Q} \approx
8 \pi \left(\frac{N_{Q}}{45}\right)^{1/2}
\frac{x_{Q}}{y_{Q}}\left(\frac{M}{M_{P}}\right)^2\cdot
\label{eq:qa}
\end{equation}
Here $N_{Q}$ is the number of e-foldings suffered by our
present horizon scale during inflation and $y_{Q}=
x_{Q}(1-7/(12x_{Q}^2)+\cdots)$ with $x_{Q} = S_{Q}/M$,
$S_{Q}$ being the value of the scalar field $S$ when the
scale which evolved to the present horizon size crossed
outside the de Sitter (inflationary) horizon. Also, from
Eq.(\ref{eq:veff}), one finds
\begin{equation}
\kappa \approx \frac{8\pi^{3/2}}{\sqrt{N_{Q}}}
~ y_{Q}~\frac{M}{M_{P}}~\cdot
\label{eq:kappa}
\end{equation}

\par
Inflation ends as $S$ approaches $S_{c}$.
Writing $S=xS_{c}$, $x=1$ corresponds to the phase
transition from $G$ to $G_{S}$ which, as it turns out,
more or less coincides with the end of the inflationary
phase (this is checked by noting the amplitude of the
quantities $\epsilon$ and $\eta$ in Eq.(\ref{eq:src})).
Indeed, the $50-60$ e-foldings needed for the inflationary
scenario can be realized even with small values of $x_{Q}$.
For definiteness, we take $x_{Q}\approx 2$. From
COBE \cite{cobe} one then obtains $M \approx 5.5 \times
10^{15}~$GeV and $\kappa \approx 4.5 \times 10^{-3}$
for $N_{Q} \approx 56$. Moreover, the primordial
density fluctuation `spectral index' $n \simeq 0.98$. We
see that the relevant part of inflation takes place at
$S \sim 10^{16}$ GeV. An important consequence of this
is \cite{lyth,lss,sugra} that the supergravity corrections
can be brought under control so as to leave inflation intact.

\par
After the end of inflation the system falls towards the
supersymmetric minima, oscillates about them and eventually
decays `reheating' the universe. The oscillating system
(inflaton) consists of the two complex scalar fields $S$ and
$\theta=(\delta \bar{\phi} + \delta\phi)/\sqrt{2}$,
where $\delta \bar{\phi}=\bar{\phi}-M$, $\delta \phi=
\phi-M$, with mass $m_{infl}=\sqrt{2}\kappa M$.

\par
In conclusion, it is important to note that the superpotential
$W$ in Eq.(\ref{eq:superpot}) leads to the hybrid inflationary
scenario in a `natural'
way. This means that a) there is no need of extremely small
coupling constants, b) $W$ is the most general renormalizable
superpotential which is allowed by the gauge and R- symmetries,
c) supersymmetry guarantees that the radiative corrections do
not invalidate inflation, but rather provide a slope along the
inflationary trajectory which drives the inflaton towards the
supersymmetric vacua, and d) supergravity corrections can be
negligible leaving inflation intact.

\acknowledgments

\par
This work is supported by E.U. under TMR contract
No. ERBFMRX--CT96--0090.

\end{document}